\documentclass{emulateapj}
\usepackage{epsf}
\usepackage{apjfonts} 
\usepackage{xspace} 
\usepackage{amsmath}
\def\figdir{.}

\def\HI{{\rm H\,I}}
\def\HII{{\rm H\,II}}
\def\GI{{\rm He\,I}}
\def\GII{{\rm He\,II}}

\def\Msun{\, M_{\odot}}

\def\fesc{f_{\rm esc}}
\def\fescg{f_{\rm esc, \HI}}

\def\dim#1{\mbox{\,#1}}
\def\figname#1#2{\figdir/#1}
\def\note#1{}
\def\hide#1{}

\begin{document}

%=================================================
\title{Escape of Ionizing Radiation from High Redshift Galaxies}
%=================================================

\author{Nickolay Y.\ Gnedin\altaffilmark{1,2,3}, Andrey
  V. Kravtsov\altaffilmark{2,3,4}, and Hsiao-Wen Chen\altaffilmark{3}}
  \altaffiltext{1}{Particle Astrophysics Center, Fermi National
  Accelerator Laboratory, Batavia, IL 60510, USA; gnedin@fnal.gov}
  \altaffiltext{2}{Kavli Institute for Cosmological Physics, The
  University of Chicago, Chicago, IL 60637, USA}
  \altaffiltext{3}{Department of Astronomy \& Astrophysics, The
  University of Chicago, Chicago, IL 60637 USA}
  \altaffiltext{4}{Enrico Fermi Institute, The University of Chicago,
  Chicago, IL 60637, USA}

\begin{abstract}
We model the escape of ionizing radiation from high-redshift galaxies
using high-resolution Adaptive Mesh Refinement
$N$-body$+$hydrodynamics simulations. Our simulations include
time-dependent and spatially-resolved transfer of ionizing radiation
in three dimensions, including effects of dust absorption.  For
galaxies of total mass $M\gtrsim 10^{11}{\ \rm M_{\odot}}$ and star
formation rates ${\rm SFR}\approx 1-5{\rm\ M_{\odot}\,yr^{-1}}$, we
find angular averaged escape fractions of $1-3$\% over the entire
redshift interval studied ($3<z<9$).  In addition, we find that the
escape fraction varies by more than an order of magnitude along
different lines-of-sight within individual galaxies, from the largest
values near galactic poles to the smallest along the galactic disk.
The escape fraction declines steeply at lower masses and SFR. We show
that the low values of escape fractions are due to a small fraction of
young stars located just outside the edge of HI disk. This fraction,
and hence the escape fraction, is progressively smaller in disks of
smaller galaxies because their HI disks are thicker and more extended
relative to the distribution of young stars compared to massive
galaxies.  Our results suggest that high-redshift galaxies are
inefficient in releasing ionizing radiation produced by young stars
into the intergalactic medium.  We compare our predicted escape
fraction of ionizing photons with previous results, and find a general
agreement with both other simulation results and available direct
detection measurements at $z\sim3$. We also compare our simulations
with a novel method to estimate the escape fraction in galaxies from
the observed distribution of neutral hydrogen column densities along
the lines of sights to long duration $\gamma$-ray bursts. Using this
method we find escape fractions of the GRB host galaxies of
$2-3$\%, consistent with our theoretical predictions.
\end{abstract}


%----------------------
\section{Introduction}
\label{sec:intro}
%----------------------

Star forming galaxies have a number of important effects on the
surrounding intergalactic medium (IGM) and subsequent gas accretion.
The ionizing radiation from galaxies is thought to be responsible for the
re-ionization of the universe \citep[e.g.,][]{mhr99,bhvs05}, altering
the thermal state of the IGM
\citep{gh98,g00c,rgs00,stre00,mmrs01,fr07}, reducing gas accretion
onto small dwarf galaxies \citep{e92,tw96,dhrw04}, and evaporating the
existing gas in small halos \citep{bl99,shaviv_dekel03,sir04}.

The amount of radiation emitted by galaxies into the IGM depends not
only on the abundance of hot, young stars, but also on the spatial
distribution of absorbing gas and 
dust in individual galaxies and their immediate surroundings.  The
escape of ionizing radiation from galaxies is therefore the focus of a
number of observational and theoretical studies.  Nevertheless, the
``escape fraction'', $\fesc$, that characterizes the fraction of total
ionizing radiation released into the IGM from individual galaxies
remains poorly constrained.  

Recent empirical measurements of the escape fraction from normal
galaxies in the local universe \citep{lfhl95,hsml01,bzaa06} and at
high redshifts \citep{gcdf02,flc03,sspa06}\footnote{Note that both
\citet{gcdf02} and \citet{sspa06} quote larger values, up to
15\%. However, the quoted numbers are for {\it relative\/} escape
fractions between the $1500\AA$ and the Lyman limit. The {\it
absolute\/} escape fractions are actually about 3\%, as we discuss in
more detail in \S \ref{sec:discussion}.}  have generally produced
modest values in the range of a few percent.  In contrast, theoretical
studies of the escape of ionizing radiation from high-redshift
galaxies have largely been inconclusive.  Many of the previous studies
have applied simplified analytic models
\citep{ds94,hl97,dsf00,rs00,wl00,co02,fmma03} that predicted a wide
range of values for $\fesc$.

A more reliable estimate can be derived from
self-consistent cosmological numerical simulations of galaxy formation
that include three-dimensional radiative transfer and model the
three-dimensional distribution of absorbing gas in and around
individual galaxies. This approach is computationally challenging and
has been attempted only recently at a range of redshifts, from
$z\sim20$ \citep{abs06} to more modest redshifts $z\ga2$
\citep{rs06,rs07}, where a comparison with the observations is
possible.

In this paper, we continue this line of work using fully
self-consistent cosmological simulations that include radiative
transfer and resolve the interstellar medium in modest-sized galaxies
at high redshifts. Our work is similar but complementary to the work
of \citet{rs06,rs07}. In both approaches, self-consistent cosmological
simulations of normal galaxies are used. However, while
\citet{rs06,rs07} use the Smooth Particle Hydrodynamics (SPH) for
modeling galaxies with modest numbers of stellar particles (up to
several thousand), we apply Adaptive Mesh Refinement (AMR) method in
this study to reach a larger spatial dynamic range. \citet{rs06,rs07}
use the exact ray-tracing method for modeling the radiative transfer
of ionizing radiation from simulated galaxies without taking into
account effects of radiative transfer during simulation, while we use
an approximate Optically Thin Variable Eddington Tensor (OTVET) method
of \citet{ga01} with radiative transfer calculations performed during
the course of simulation self-consistently.

Most importantly, in this work we explicitly include absorption by
dust, which allows a direct comparison with 
observational estimates of the escape fraction.  Given the differences
between \citet{rs06,rs07} and our approaches, a comparison
of the results 
is useful for estimating the range of systematic numerical
uncertainties in the problem, where a rigorous convergence study is
not yet feasible.

The paper is organized as follows. In \S~\ref{sec:sims} we describe
the simulations and numerical techniques used in our study. 
In \S~\ref{sec:method} we provide operational definition
of the escape fraction and in \S~\ref{sec:dust} we describe our
model for dust absorption of the ionizing radiation. We present
the results of the study in \S~\ref{sec:results} and discuss
their implications in \S~\ref{sec:discussion}.

%---------------------
\section{Simulation}
\label{sec:sims}
%---------------------

The simulation we use in this paper is described in detail by
\citet[][where it is labeled as FNEQ-RT]{tkg07}. Here we briefly
summarize its setup and numerical parameters.

The simulation is performed using the Eulerian, gas dynamics +
$N$-body Adaptive Refinement Tree (ART) code \citep{kkk97,kkh02}. A
large dynamic range is achieved through the use of Adaptive Mesh
Refinement (AMR) in both the gas dynamics and gravity calculations.
The calculation is started from a random realization of a Gaussian density
field at $z=50$ in a periodic box of $6h^{-1}\ \rm Mpc$ in a flat
$\Lambda$CDM model ($\Omega_m=1-\Omega_{\Lambda}=0.3$,
$\Omega_b=0.043$, $h=0.7$, $n_s=1$, and $\sigma_8=0.9$). A Lagrangian
region corresponding to five virial radii of a Milky Way sized galaxy
at $z=0$ is identified and evolved with $2.6\times 10^6$ dark matter
particles with masses of $9.2\times 10^5h^{-1}\Msun$.

The code employs a uniform $64^3$ grid to cover the entire computational
box. The Lagrangian region is, however, always unconditionally refined
to the third refinement level, corresponding to the effective grid
size of $512^3$. As the matter distribution evolves, the code
adaptively and recursively refines the mesh in high-density regions
beyond the third level up to the maximum allowed 9th refinement level,
which corresponds to the comoving spatial resolution of $260\dim{pc}$,
or physical resolution of $65 (50) \dim{pc}$  at $z=3 (4)$.

The ART code computes metallicity-dependent, non-equilibrium gas
cooling ``on the fly'' based on the abundances of five atomic species
and molecular hydrogen.  Star formation and feedback (both radiative
and thermal via supernova explosions and stellar winds) are included,
as described in \citet{tkg07}.

The self-consistent 3D radiative transfer of UV radiation from
individual stellar particles is followed with the OTVET algorithm
\citep{ga01}. The details of our implementation of the OTVET algorithm
on adaptively refined meshes will be described elsewhere (Gnedin \&
Kravtsov 2007, in preparation). A comparison of our radiative transfer
scheme with several other existing numerical implementations for the
time-dependent and spatially-inhomogeneous radiative transfer is
presented in \citet{icam06} and in the follow-up paper (Iliev et al.\
2007, in preparation).

In addition to the main progenitor of the Milky Way type galaxy, the
fully refined Lagrangian region contains several dozens of smaller
galaxies spanning a wide range of masses. We will use all these
systems to estimate the escape fraction of ionizing radiation from
star forming regions for galaxies of different masses and star
formation rate (SFR).

%---------------------
\section{Method}
\label{sec:method}
%---------------------

Because our simulation includes an approximate treatment of the
radiative transfer self-consistently, the non-equilibrium abundances
of all ions for each cell in the AMR grid hierarchy are available
throughout the calculation. We can thus measure the opacity at a given
frequency along a given direction from any point in the simulation
volume to any other point.

Since the opacity can only increase along a given line of sight, the
escape fraction is, in general, a function of the distance from the
source.  We choose to measure this distance in units of the virial
radius of a given galaxy and we measure the escape fraction at 0.5, 1,
and 2 virial radii from the center of the galaxy (defined as the peak
of the dark matter density). Unless specified otherwise, hereafter we
use the escape fraction at one virial radius as our fiducial working
definition of the escape fraction.

In computing the escape fraction, it is important to separate
satellite galaxies from isolated ones.  When a small satellite galaxy
is located within a larger galaxy, the concept of virial radius and
escape fraction becomes ambiguous.  Therefore, we only consider
isolated galaxies and only include sources that actually reside in
(i.e.\ gravitationally bound to ) the main galaxy, thus excluding the
satellites. In computing the continuum 
absorption by ionic species we include the detailed velocity structure
of the galactic and infalling gas, although we find that ignoring the
gas velocity field makes no measurable difference on the escape
fraction.

Operationally, the escape fraction $\fesc$ of a specific galaxy at a given
redshift is then a function of two variables: the frequency and the
direction of propagation of escaping radiation:
\begin{equation}
\fesc \equiv \fesc(\nu,\vec{\theta}).
\end{equation}
For studies of reionization of the universe and the Lyman-alpha forest,
the most important quantity is the escape fraction of
ionizing radiation,
\begin{equation}
\fesc^{j}(\vec{\theta}) \equiv 
  \frac{\int d\nu\,\fesc(\nu,\vec{\theta})\sigma^{j}_\nu S_\nu}
  {\int d\nu\,\sigma^{j}_\nu S_\nu},
  \label{eq:fion}
\end{equation}
where $\sigma^{j}_\nu$ is the photo-ionization cross-section for the
specie $j$ ($j=\HI,\GI,\GII$) and $S_\nu$ is the spectrum of
the sources of radiation. In our simulation, we include only stellar
sources and compute the ionizing radiation spectra using the
Starburst99 package \citep{lsgd99}.  We assume continuous star
formation, 0.04 solar metallicity (typical of galaxies in our
simulation), and a Salpeter initial mass function over a mass range
from 1 to 100 M$_\odot$. The spectral shape is shown in Figure~4 of
\citet{rgs02a}. Note that the Starburst99 spectral shape is computed
for the unobscured stellar population, which is the appropriate
spectral shape to use in equation (\ref{eq:fion}), which explicitly
includes the $\fesc$ factor.

However, observationally the escape fraction of ionizing radiation is
difficult to determine, since it requires measuring the whole ionizing
spectrum. Instead, the escape fraction at the hydrogen ionization edge
(Lyman limit), $\fesc(\nu_0)$, is usually measured in observational
studies. We present the relationship between these two quantities in
the appendix.

%---------------------
\section{Absorption by Dust}
\label{sec:dust}
%---------------------

Incorporating dust absorption into the simulations for calculating the
escape fraction is critical, because dust may contribute significantly
to the absorption of ionizing radiation
\cite[e.g.][]{wd01}. Unfortunately, properties of dust in
high-redshift galaxies are not known well. In particular, dust
absorption cross section depends on the dust composition and grain
size distribution, which is measured only in nearby galaxies. In our
analysis, we adopt the dust extinction curves for Large and
Small Magellanic Clouds (LMC and SMC respectively) from \citet{wd01}
as representative of high redshift galaxies,
because low metallicities of these galaxies are closer to typical
metallicities of high-redshift galaxies in our simulation.

A convenient parameterization for the dust extinction law in
LMC and SMC is given by \citet{p92}, based on the earlier data,
\begin{equation}
  \sigma^{\rm d}_\nu = \sigma_0 \sum_{i=1}^7
  f(\lambda/\lambda_i,a_i,b_i,p_i,q_i), 
  \label{eq:pei}
\end{equation}
where the fitting function $f$ has the form
\begin{equation}
  f(x,a,b,p,q) = \frac{a}{x^p + x^{-q} + b}.
  \label{eq:fit}
\end{equation}

\begin{deluxetable}{lccccc}
\tablecaption{Parameters for the dust extinction law fits\label{tab:pei}}
\tablewidth{\columnwidth}
\tablehead{
\colhead{Term} & \colhead{$\Lambda_i$ ($\mu$m)} &
\colhead{$a_i$} & \colhead{$b_i$} & \colhead{$p_i$} & \colhead{$q_i$}
}
\startdata\\
\cutinhead{SMC dust model}
1 & 0.042 & 185   & 90    & 2 & 2\\
2 & 0.08  & 27    & {\bf 15.5}\tablenotemark{a} &  4 & 4\\
3 & 0.22  & 0.005 & -1.95 & 2 & 2\\
4 & 9.7   & 0.010 & -1.95 & 2 & 2\\
5 & 18    & 0.012 & -1.8  & 2 & 2\\
6 & 25    & 0.030 & 0     & 2 & 2\\
7 & {\bf 0.067} & {\bf 10} & {\bf 1.9} & {\bf 4} & {\bf 15}\\
\cutinhead{LMC dust model}
1 & 0.046 & {\bf 90} &  90    & 2 & 2\\
2 & 0.08  & 19    & {\bf 21} & 4.5 & 4.5\\
3 & 0.22  & 0.023 &  -1.95 & 2 & 2\\
4 & 9.7   & 0.005 &  -1.95 & 2 & 2\\
5 & 18    & 0.006 &  -1.8  & 2 & 2\\
6 & 25    & 0.02  &   0    & 2 & 2\\
7 & {\bf 0.067} & {\bf 10} & {\bf 1.9} & {\bf 4} & {\bf 15}\\
\enddata
\tablenotetext{a}{Values changed from \citet{p92} are shown in bold.}
\end{deluxetable}

We correct the fits of \citet{p92} using the newer data from
\citet{wd01} by adding the seventh term in equation (\ref{eq:pei}) in
order to account for the narrow and asymmetric shape of the FUV peak
in the dust extinction law. We also change values for some parameters from
those adopted by \citet{p92}. The values for the parameters we use are
listed in Table \ref{tab:pei}, and we show in bold changes from
\citet{p92}.

The overall normalization for the dust cross section is determined by
parameters $\sigma_{0, {\rm LMC}}=5.6\times10^{-22}\dim{cm}^2$ and
$\sigma_{0, {\rm SMC}}=1.1\times10^{-22}\dim{cm}^2$. With these
functional forms, the dust cross sections for both LMC and SMC fit the
plotted curves of \citet{wd01} to a few percent accuracy.

In order to limit the range of possible dust effects, we consider three
extreme dust models: 1) a model with no dust at all, 2) a
model in which we assume that the dust column density scales with
neutral (atomic and molecular) gas column density,
\begin{equation}
  N_{\rm dust} = \frac{Z}{Z_0} \times (N_{\HI}+2N_{\rm H_2})
  \label{eq:insub}
\end{equation}
(where $Z$ is the gas metallicity), and 3) a model where the dust
column density scales with the total gas column density (both neutral
and ionized),
\begin{equation}
  N_{\rm dust} = \frac{Z}{Z_0} \times N_{\rm H},
  \label{eq:nosub}
\end{equation}
where $Z_0=0.32$ ($0.2$) is the gas-phase metallicity of the LMC
\citep[or SMC,][]{wlbh97,wfsy99}. Note that emission line studies
\citep{ppr00,kw06} usually indicate somewhat larger values for both metallicities
than the absorption line metallicities that we adopt. 
If we adopted larger values for the LMC and SMC
metallicities, the dust effect on our measured escape fractions would
be even {\it smaller\/}.

Physically, equation (\ref{eq:insub}) implies a complete instant
sublimation of dust in the ionized gas, while equation
(\ref{eq:nosub}) implies no dust sublimation at all. Of course, the
truth lies somewhere in between, but these two extreme cases bracket
the range of possibilities.

\begin{figure}[t]
\plotone{\figname{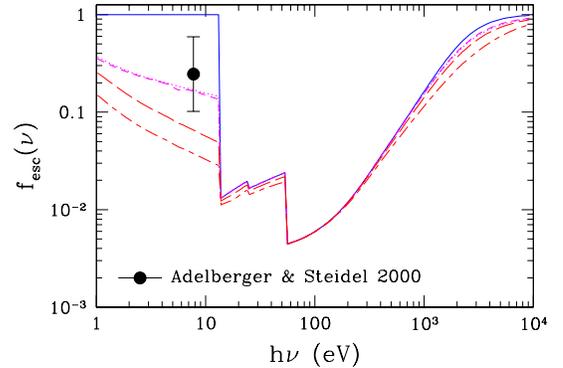}{f1.eps}}
\caption{The frequency dependence of the angular averaged escape
  fraction for the central galaxy at $z=3$ for several dust models: no
  dust (solid line), SMC dust with instant sublimation (eq.\
  [\ref{eq:insub}], dotted line), LMC dust with instant sublimation (eq.\
  [\ref{eq:insub}], short-dashed line), and SMC dust with no
  sublimation (eq.\ [\ref{eq:nosub}], long-dashed line). In addition,
  we also show with a short-long-dashed line the SMC no
  sublimation model with the dust cross section increased by a factor
  of 3, as a certain {\it overestimate\/} of the dust absorption. The
  black point with error-bars shows the extinction at $1600\AA$ from
  \citet[][derived as $A_{1600}=4.43+1.99\beta$ with
  $\beta=-1.46\pm0.48$]{as00}.}
\label{figFN}
\end{figure}
Figure~\ref{figFN} shows the angular averaged escape fraction for the
Milky Way progenitor galaxy at $z=3$ as a function of frequency. We
show several dust models, together with typical reddening correction
for Lyman Break Galaxies from \citet{as00}. We also show the SMC no
sublimation dust model with dust cross section arbitrarily increased
by a factor of 3. As we have mentioned above, the no sublimation model
is likely to overestimate the effect of dust absorption. Since it is
highly unlikely that uncertainties in our adopted value for the SMC
metallicity and the dust extinction curve are as large as a factor of
3, this latter model serves as an absolute (and, likely, implausibly
high) upper limit for the dust absorption.

\begin{figure}[t]
\plotone{\figname{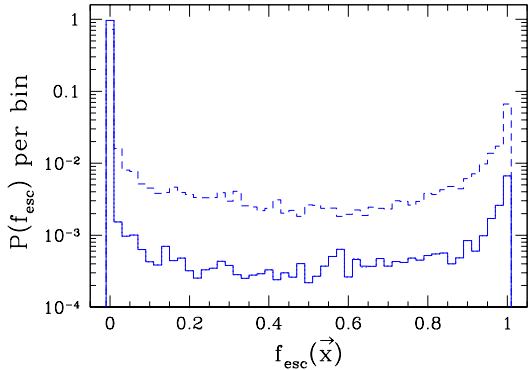}{f2.eps}}
\caption{The luminosity weighted distribution of the escape fraction
  at the Lyman limit in a random direction on the sky from all
  positions within the central galaxy for the SMC instant sublimation
  dust model (our fiducial one). Three separate distributions are
  shown: absorption by gas only (dotted line), absorption by dust only
  (dashed line), and total absorption (dust + gas, solid line). Notice
  that the gas-only distribution is essentially identical to the total
  distribution, because $Z\sigma^{\rm d} \ll \sigma^{\HI}$ at the
  Lyman limit; so the dotted line coincides with the solid
  one. The histograms are normalized to add up to 1, so the spike at
  $\fesc\approx0$ contains most of the data points.\newline}
\label{figLS}
\end{figure}
A rather unexpected feature of Figure~\ref{figFN} is that
the escape fraction above the Lyman limit is almost independent of the
dust absorption. In order to understand this phenomenon, we show in
Figure~\ref{figLS} the distributions of escape fractions from all
positions within the central galaxy in a fixed, randomly selected
direction on the sky. The
distributions have similar shapes: they are weak
functions of $\fesc$ for $0<\fesc<1$ and exhibit a primary peak at
$\fesc\approx0$ and a secondary peak at $\fesc\approx1$.  Thus, the
escaping radiation is produced {\it not by sources in a semi-opaque
medium,\/} with each source attenuated by a similar amount, but {\it
by a small fraction of essentially unobscured sources\/}. We discuss
this point in detail in \S~\ref{sec:results} below (see
Fig.~\ref{figDK}).

For a distribution like those in Figure~\ref{figLS}, the effects of
dust absorption can be easily understood, if we
ignore the ``translucent points'' at $0<\fesc<1$ and use a toy-model
distribution for the gas-only $\fescg$,
\begin{equation}
  p(\fescg) = \alpha\delta(1-\fescg) + (1-\alpha)\delta(\fescg),
  \label{eq:fdist}
\end{equation}
where $\alpha\ll1$ is constant and $\delta(x)$ is a Dirac delta
function. The average escape fraction in this model is simply
\begin{equation}
  \langle\fescg\rangle = \int_0^1 \fescg p(\fescg) d\fescg = \alpha.
  \label{eq:favg}
\end{equation}
If the dust absorption is included, then the escape
fraction at each location is changed by a factor that
depends on the dust opacity,
$$
  \fesc = \fescg e^{-\tau_{\rm d}} \equiv e^{-\tau_{\HI}-\tau_{\rm d}}.
$$
At the locations where $\fescg\approx1$ ($\tau_{\HI}\ll 1$),
the effect of dust is negligible (since $\tau_{\rm d}\ll\tau_{\HI}$).
At locations where $\tau_{\rm d}\ga 1$, the hydrogen opacity is
already so large that no radiation escapes from this position,
irrespective of how much dust is mixed with the gas. As the result,
the escape fraction does not change at all.

In reality, ``translucent points'' with $0<\fesc<1$ are affected by
dust, but their integrated contribution to the average escape fraction
remains small.  In the no sublimation dust model, the situation
may be more complex, since there is dust absorption in the ionized
gas. However, Figure~\ref{figFN} demonstrates that this is not a large
effect.

At frequencies below the Lyman limit the distribution of $\fesc$ from
dust is similar to Figure~\ref{figLS}, but there is no gas
absorption. Because dust absorption is weaker than gas absorption,
the average escape fraction at these frequencies is much larger and
increases to unity as the frequency decreases down to infra-red.

The important result of this section is that, while dust absorption is
the dominant effect for UV radiation below the Lyman Limit, it does
not substantially affect the escape fraction of ionizing radiation 
in our model galaxies, as Figure~\ref{figFN} demonstrates.

In the rest of this paper, we adopt the SMC instant sublimation model
as our fiducial dust model, but we note that our final results are not
sensitive to this particular choice.

%-------------------
\section{Results}
\label{sec:results}
%-------------------

\begin{figure}[t]
\plotone{\figname{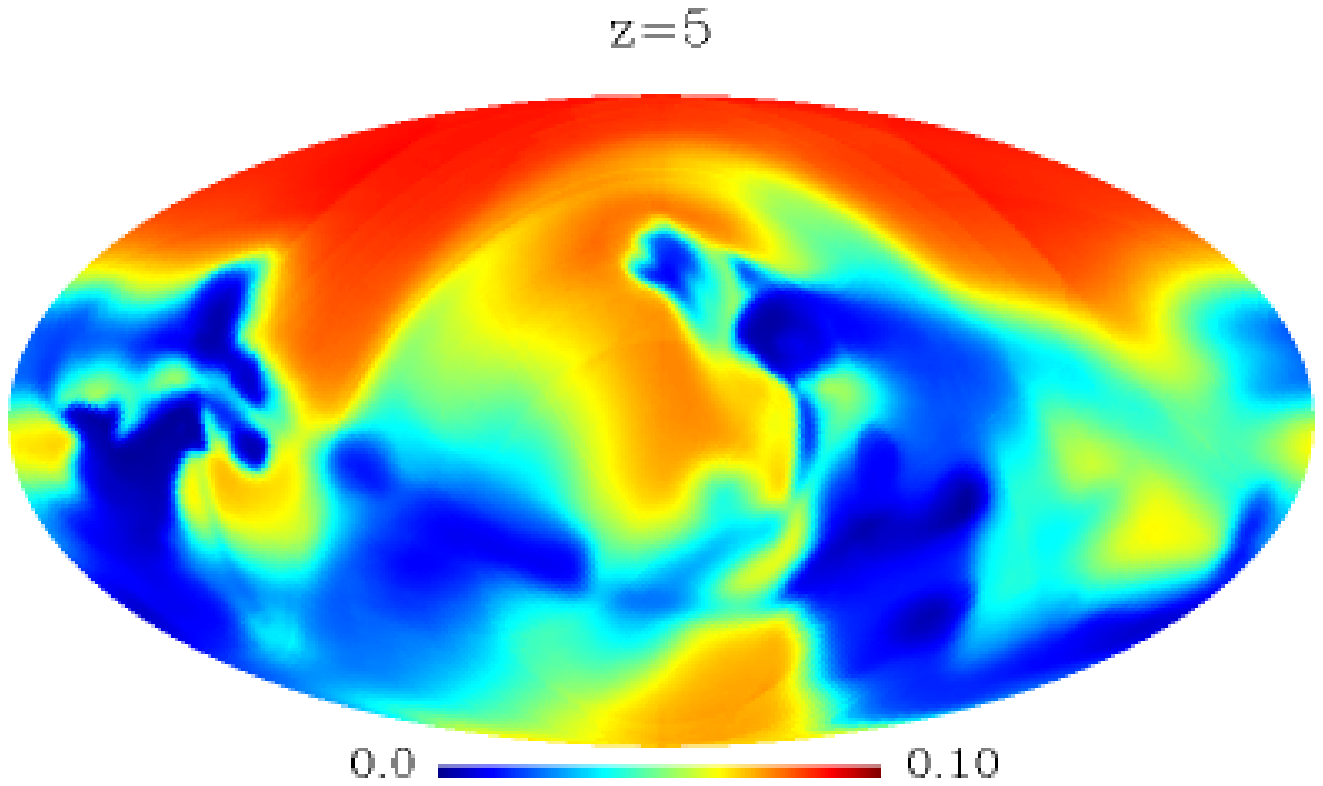}{f3a.eps}}
\plotone{\figname{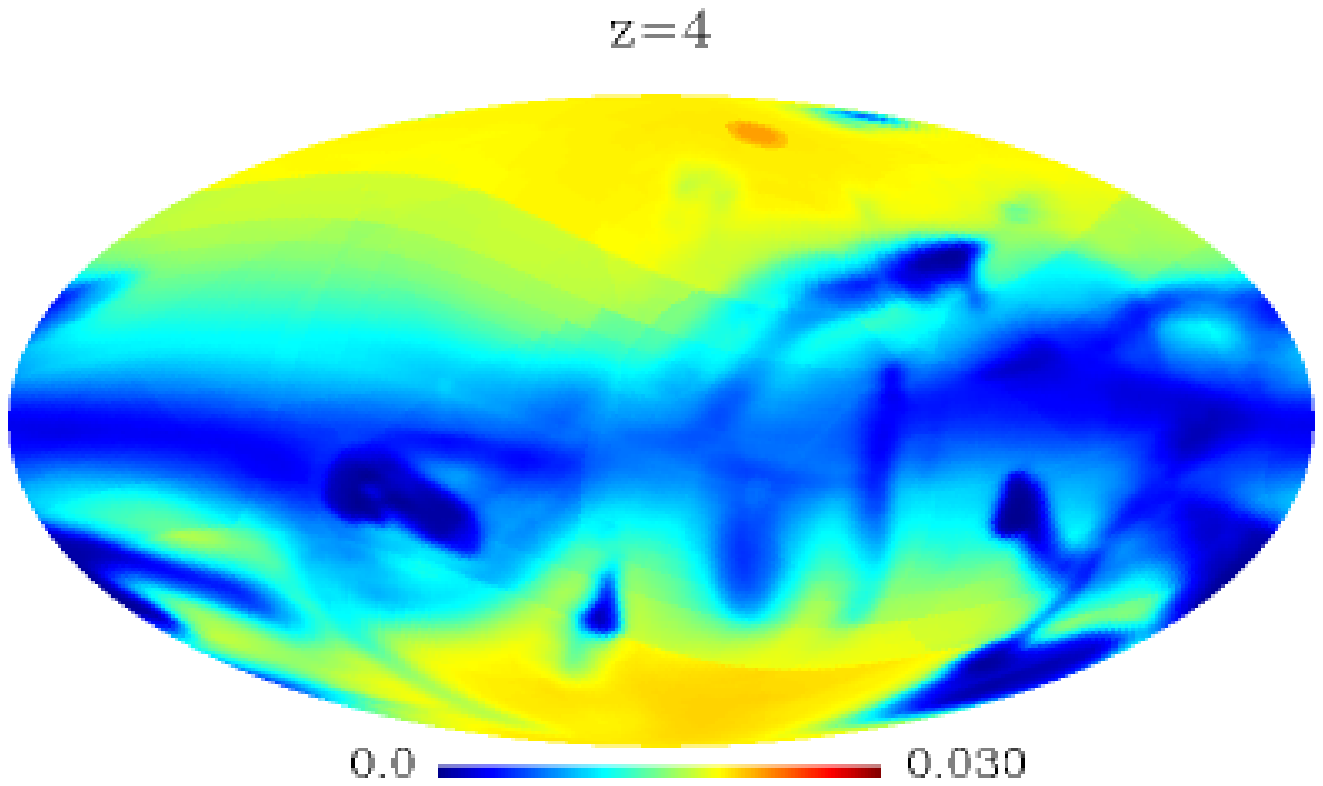}{f3b.eps}}
\plotone{\figname{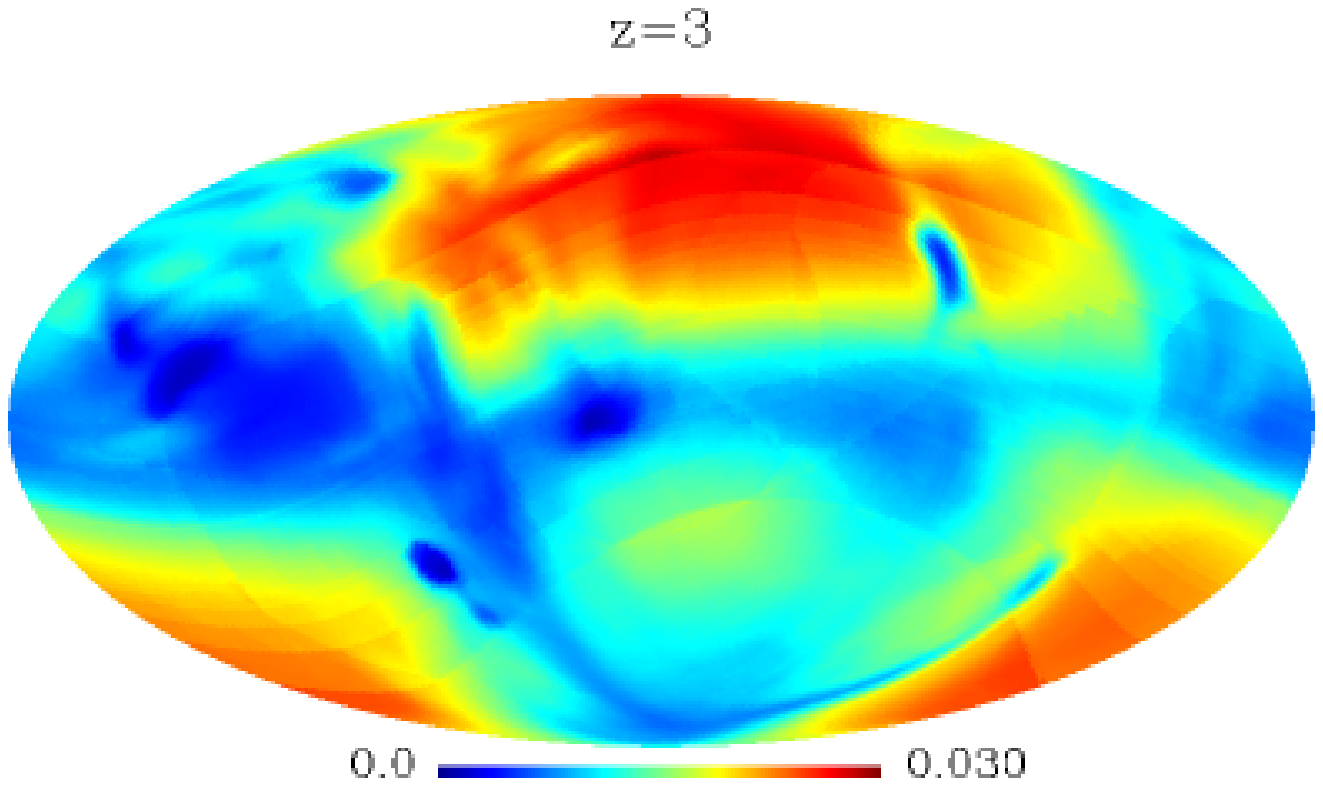}{f3c.eps}}
\caption{The escape fraction of $\HI$ ionizing radiation $\fesc^{\HI}$
  as a function of direction, as visible from the center of the Milky
  Way type galaxy, at the virial radius of the galaxy at $z=5$, $4$,
  and $3$. The sky is oriented in coordinates aligned with galactic
gas disk such that the disk is horizontal in the middle of the plot. Note a
  different color scale for $z=5$ image.} 
\label{figIM}
\end{figure}

Figure~\ref{figIM} shows the escape fraction of hydrogen ionizing
radiation $\fesc^{\HI}$ (eq.\ \ref{eq:fion}) as seen from the center
of the central galaxy (the Milky Way progenitor) at three different
redshifts. The celestial coordinates in Figure\ \ref{figIM} are
aligned with the principal axes of the galaxy (``galactic''
coordinates), with the plane of the disk of the galaxy crossing the middle of
the plot. Typically, the escape fraction is close to zero along the
plane of the disk and approaches maximum values near the poles,
although there are significant small-scale variations at all redshifts
that underscore the complex, perturbed nature of high redshift disk
galaxies. At $z=5$ the galaxy is experiencing a substantial major
merger (and a lesser one at $z=3$), so the angular distribution of
escape fractions is more irregular at these epochs.

Another interesting feature of the $\fesc$ distribution shown in
Figure~\ref{figIM} is a few small very opaque (dark blue) clouds of
gas that block ionizing radiation at larger distances.  These clouds
can be counterparts of the Lyman Limit absorbers observed in the
spectra of distant quasars. Previous studies, based on lower
resolution simulations, have indicated that Lyman Limit systems tend
to cluster around large galaxies \citep{kg07}. Our simulations
confirm that such clouds do exist around high-redshift galaxies.

We have checked that, as we integrate further in distance, more of
Lyman Limit systems fall inside the radius of integration, and the
``sky'', as seen from the center of the galaxy, appears progressively
more opaque.  The sky should become completely opaque at a distance of
a few mean free paths for ionizing radiation. At $z=4$ the mean free
path for ionizing radiation is about $85\,h^{-1}$ co-moving Mpc
\citep{m03}, much larger than the size of our computational box. Thus,
we cannot actually reach the full opacity with our current
simulation. However, the important point to make is that the mean free
path is much larger than a virial radius of any galaxy at these
redshifts, so the concept of the ``escape fraction'' is well defined
and robust at intermediate redshifts.

\begin{figure}[t]
\plotone{\figname{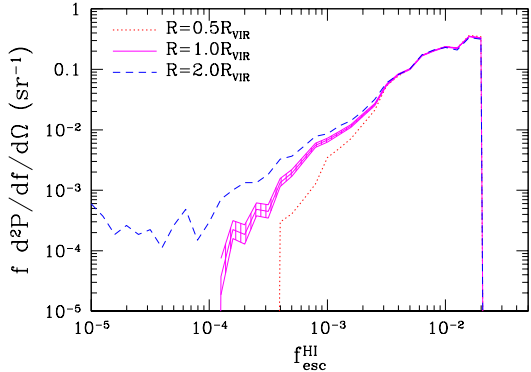}{f4.eps}}
\caption{The probability per unit log and per unit solid angle
  for having a specific 
  value of the escape fraction of $\HI$ ionizing radiation for the
  central galaxy at $z=4$ for 3 distances from the galaxy center (as
  measured in units of the galaxy virial radius). The shaded region
  around the middle curve shows the Poisson errors due to the finite
  number of pixels in the angular maps.} 
\label{figHA}
\end{figure}

Figure~\ref{figHA} illustrates this point in a quantitative way. It
shows the probability density for having a specific value of the
escape fraction on a celestial sphere at different distances from the
center of the main progenitor of the Milky Way-sized galaxy. As the
distance form the galaxy increases, the tail of the distribution at
low escape fractions grows, as Lyman limit systems cover a
progressively larger fraction of the sky. However, at distance
comparable to the virial radius of the galaxy the effect of Lyman
limit systems is small, as can be seen from the plot: the fraction of
the sky below $\fesc^{\HI}=0.003$ increases from 3\% to only 6\% as
the distance increases from $0.5R_{\rm VIR}$ to $2R_{\rm VIR}$.

Poisson errors on the probability density shown in Figure\ \ref{figHA}
remain small for escape fractions well below $10^{-3}$, indicating
that angular maps from Figure\ \ref{figIM} resolve the structure in the
gas down to that level. Of course, the small-scale structure in the
gas is limited by the finite resolution of our simulation.

\begin{figure}[t]
\plotone{\figname{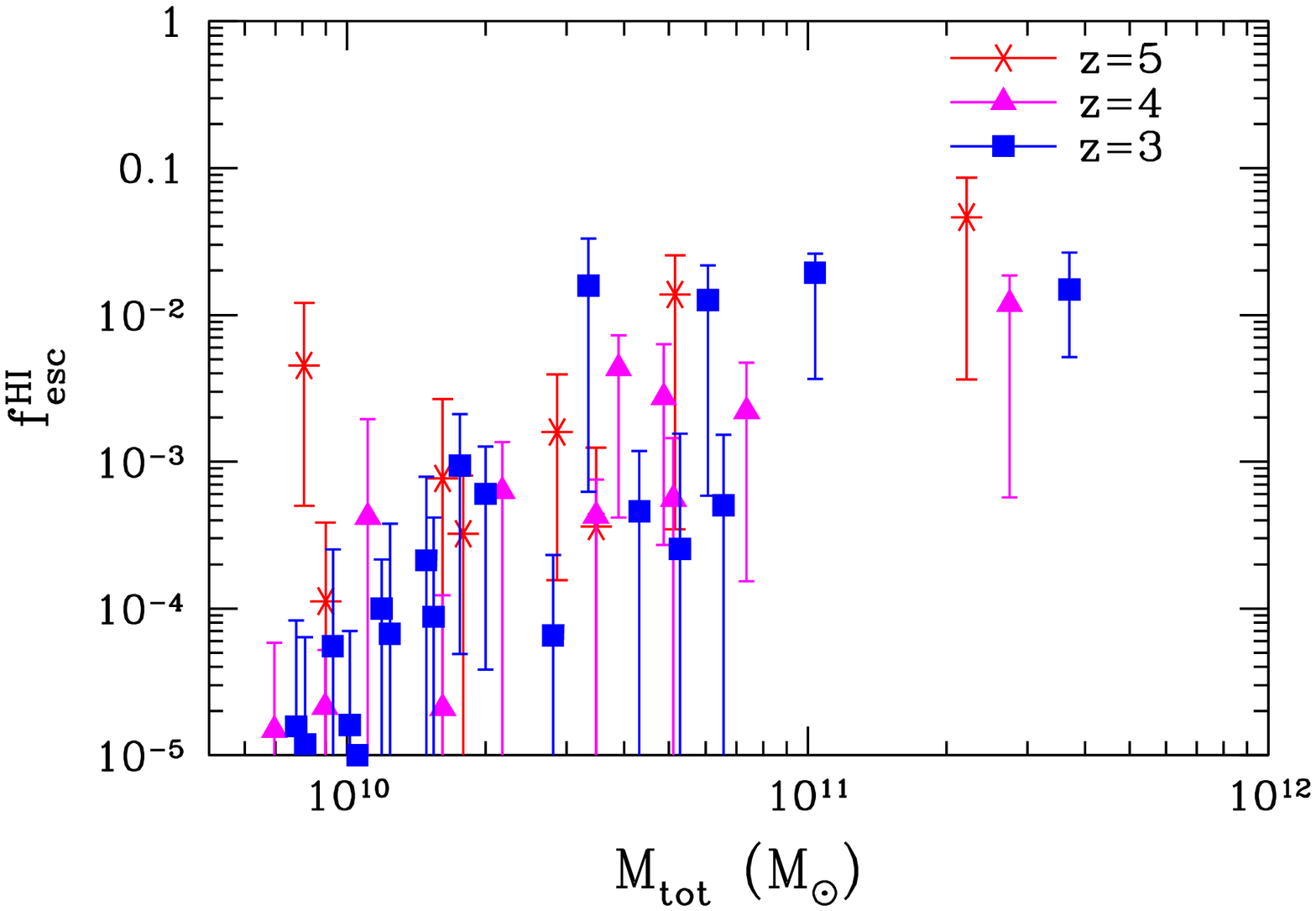}{f5a.eps}}
\plotone{\figname{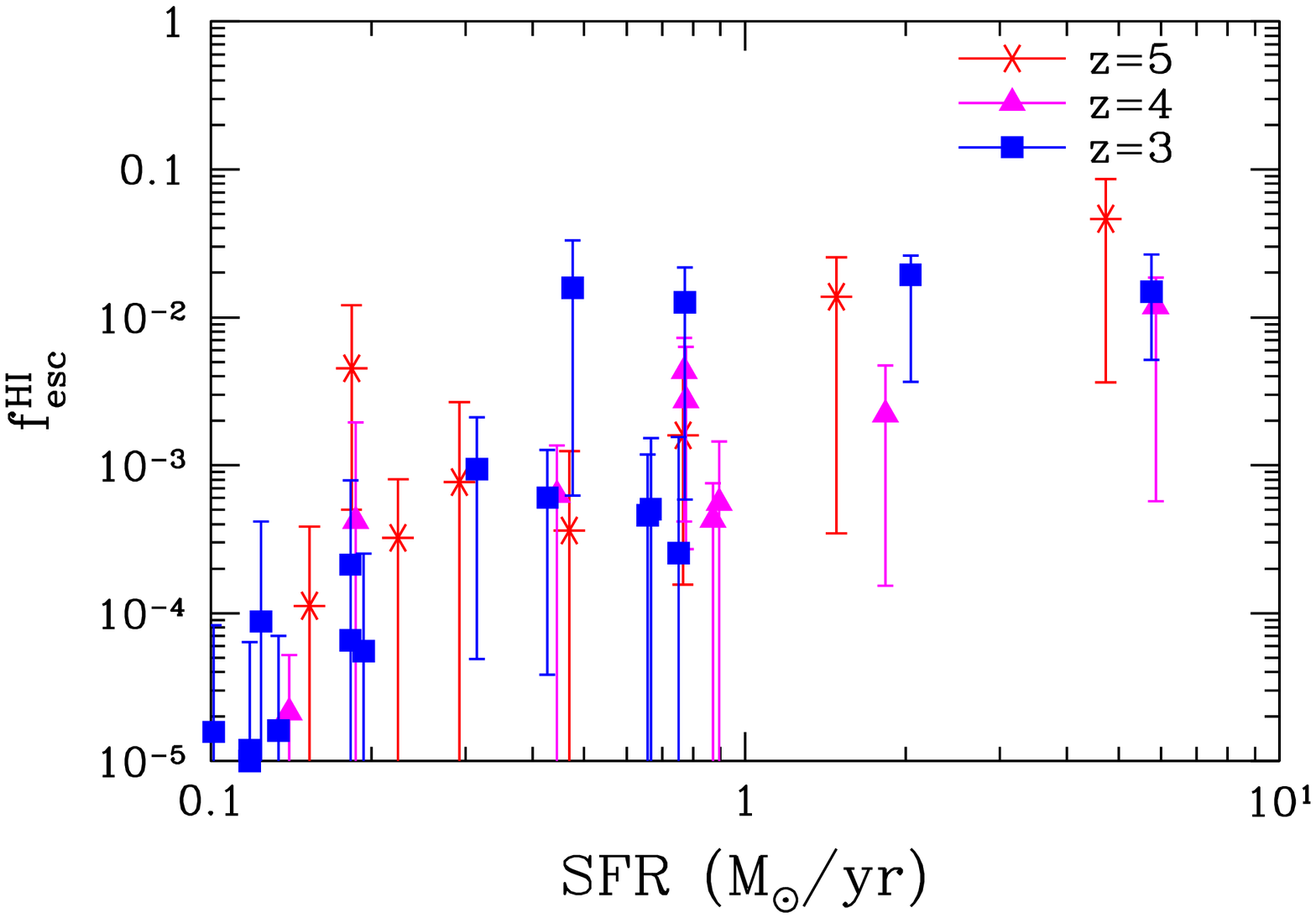}{f5b.eps}}
\caption{The angular averaged escape fraction of $\HI$ ionizing
  radiation as a function of galaxy mass (top) and star
  formation rate 
  (bottom) at three redshifts (squares). The error-bars show the
  10\%--90\% range for all possible directions.}
\label{figFM}
\end{figure}

Figure~\ref{figFM} presents the main result of this paper: the angular
averaged escape fraction for hydrogen ionizing radiation as a function
of galaxy mass or star formation rate at a range of redshifts. We only
show results for the galaxies which are resolved down to the maximum
ninth level of refinement. This resolution criterion corresponds
approximately to a minimum mass of $10^{10}\Msun$ (or $\gtrsim 10^4$
dark matter particles).  A general trend of increasing escape fraction
with increasing galaxy mass and SFR is clearly observed: the escape
fraction changes by approximately two orders of magnitude from
$\approx 10^{-4}$ to $\approx 0.02-0.05$ for total masses between
$10^{10}$ to $4\times 10^{11}\Msun$ (or SFR from $0.1$ to
$\approx 7\,\Msun\dim{yr}^{-1}$). At the same time, the figure also
shows that there is little change with redshift, from $z=5$ to $z=3$,
either in the values of the escape fraction or in the dominant trend
with mass or SFR.

\begin{figure}[t]
\plotone{\figname{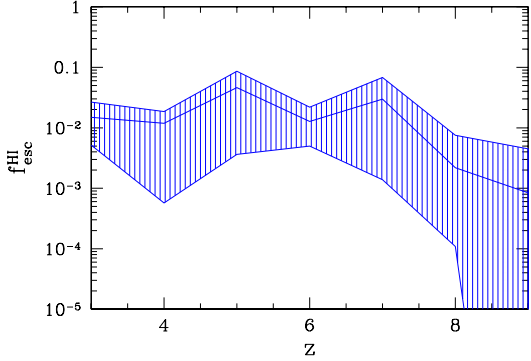}{f6.eps}}
\caption{The evolution of the angular averaged escape fraction of
  $\HI$ ionizing radiation for the most galaxy ($M_{\rm
  tot}=4\times10^{11}\Msun$ at $z=3$). The shaded band shows the
  10\%--90\% range for all possible directions.}
\label{figFZ}
\end{figure}

Figure~\ref{figFZ} further illustrates the lack of redshift evolution in
the escape fraction found in our simulations. The figure shows the 
average escape fraction of the most massive Milky Way progenitor in the
simulation at different redshifts, along with the variation in escape fraction
in different directions. While the escape fraction fluctuates with
redshift, a spread in the escape fraction in different directions at a
given redshift is much larger than the variation in the average escape
fraction with redshifts at $z<7$. A similarly weak trend is found
by most well resolved galaxies from our simulation.

\begin{figure}[t]
\plotone{\figname{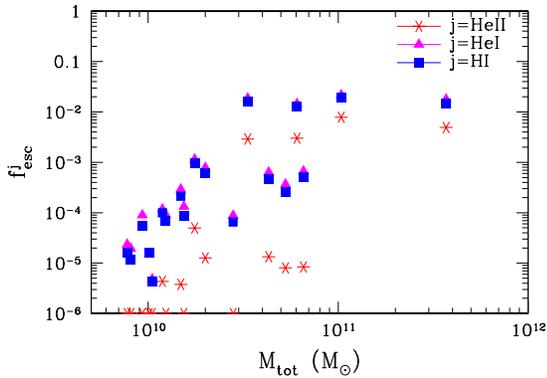}{f7.eps}}
\caption{The mass dependence of the escape fraction of ionized
  radiation for three ionic species at $z=3$ (similar to the top panel
  of Figure\ \ref{figFM}). We show only well-resolved galaxies and omit
  error-bars for clarity. The escape fractions are clumped at the
  level of $10^{-6}$ to allow all points to be visible on the graph.}
\label{figFI}
\end{figure}

There is clearly a significant drop in the escape fraction at $\GII$
ionization threshold. This is further illustrated in Figure\
\ref{figFI}, which shows all three escape fractions for ionizing
radiation as a function of the total mass for model galaxies. We note
that $\HI$ and $\GI$ are invariably close to each other, while $\GII$
escape fractions are systematically much lower. This behavior is again
completely expected, as only active galactic nuclei are thought to
fully (i.e.\ doubly) ionize helium.

\begin{figure}[t]
\plotone{\figname{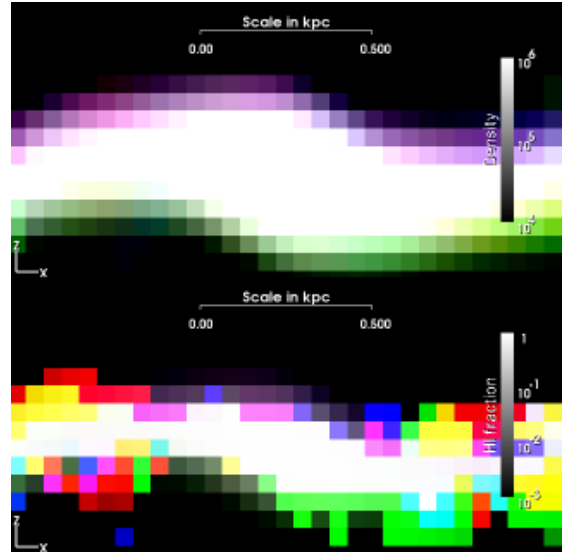}{f8.eps}}
\caption{The pseudo-color composite edge-on view of the main
  galaxy. The top half shows the gas density, while the bottom half
  depicts the neutral hydrogen fraction. Each image is composed of 3
  snapshots from the simulation taken $10\dim{Myr}$ apart at $z\sim4$
  and combined as red, green, and blue channels of a color image. The
  predominance of purple (red and blue) color on top and green color
  on bottom indicates that the galactic disk oscillates globally on a
  time scale of about $20\dim{Myr}$. The oscillation pattern is much
  more chaotic in $\HI$ image outside of about $1\dim{kpc}$. (This
  figure should be viewed in color.)}
\label{figDK}
\end{figure}

We also study the mechanism for the escape of ionizing radiation from
galactic disks in our simulations and for the predicted small values
of $\fesc$. As Figure~\ref{figLS} demonstrates, the ionizing radiation
that escapes is emitted preferentially by unobscured sources, with
only a small fraction of all sources being unobscured at
any given time (rather than being emitted by the majority of sources,
which are only partially obscured). In order to visualize the regions
of a galaxy where these unobscured sources reside,
Figure~\ref{figDK} shows a pseudo-color composite image of three
different snapshots from the simulation closely spaced in time.

The edge-on disk of the main galaxy is shown in the total gas density
(top) and in the $\HI$ fraction (bottom). Three snapshots
$10\dim{Myr}$ apart at $z\sim4$ are overlaid together so that the
first snapshot is shown in red, the second in green, and the last one
in blue. If the gas distribution did not change between the snapshots, then the
image would appear as an equal combination of red, green, and blue,
i.e.\ as a pure grayscale. In reality, however, the upper surface of
the disk appears purple (red plus blue), while the bottom surface is
pure green. This means that the galactic disk oscillates with a period
of about $20\dim{Myr}$: the disk bended downwards between the first
(red) snapshot and the second (green) snapshot, but returned to the
first configuration in the third (blue) snapshot, so that the first
and third snapshots look almost identical, blending into a single
purple color. Oscillating modes with shorter wavelengths are also
visible in the image.

This behavior is clearly visible in the total gas density image and in
the $\HI$ fraction image within the inner $0.5-0.7\dim{kpc}$; the
$\HI$ disk oscillates much more violently at larger radii. This is not
surprising, because the oscillations of the $\HI$ edge are subject to
the positive feedback.  As the $\HI$ edge oscillates and uncovers
ionizing sources, the ionizing intensity in a given point is likely to
increase (similarly to how the ionizing intensity increases when
isolated $\HII$ regions merge during reionization).

Figure~\ref{figDK} thus shows why ionizing radiation of some young
stars can leave the galaxy with virtually no attenuation. The cold
$\HI$ disk, where young stars form is very thin ($\sim
100-200\dim{pc}$). Some of the stars may have sufficient velocity to move
outside the edge of the $\HI$ disk while they are still young, bright
emitters of UV photons. For example, a star traveling at 
$10\dim{km}\dim{s}^{-1}$ will move by $\approx 100\dim{pc}$ in $10^7$ years.
In addition, the outer edge of the $\HI$ disk is changing significantly
on the same time scale. These changes can expose some of the young
stars and clear the way for the ionizing radiation to leave the
system. {\em The escaped ionizing radiation is thus mostly due to a
small fraction of stars in a thin shell surrounding the $\HI$ disk.}

The relative constancy of the escape fraction with redshift or SFR
that we find is then a simple consequence of the galactic disks
oscillating by a comparable amount on $10\dim{Myr}$ time scale.  Note
that this time scale is close to the local dynamical time scale of the
dense high-redshift disks and lifetime of massive, UV-bright
stars. Such oscillations are thus likely dynamical in origin: the
disk can be perturbed by mergers and interactions with satellites, which
are only weakly correlated with SFR or redshift (unless a disk goes
into a starburst phase).

This picture also explains why the escape fraction decreases steeply
for lower mass galaxies. Gas distribution in dwarf galaxies is
relatively more extended in comparison to the stellar distribution in
our simulations, due to lower efficiency of star formation in dwarfs
relative to massive galaxies \citep[see][for detailed
discussion]{tkg07}.  Stars are thus concentrated towards central
regions of the disk and a smaller fraction of them can reach the edge
of $\HI$ disk or get exposed by fluctuations of its boundary.
Note, however, that resolution of dwarf galaxies in our simulations
is considerably worse than for the massive galaxies. This explanation
and the trend with mass will thus need to be
verified in the future by higher resolution simulations.

%-----------------------------------
\section{Discussion and Conclusions}
\label{sec:discussion}
%-----------------------------------

\begin{figure}[t]
\plotone{\figname{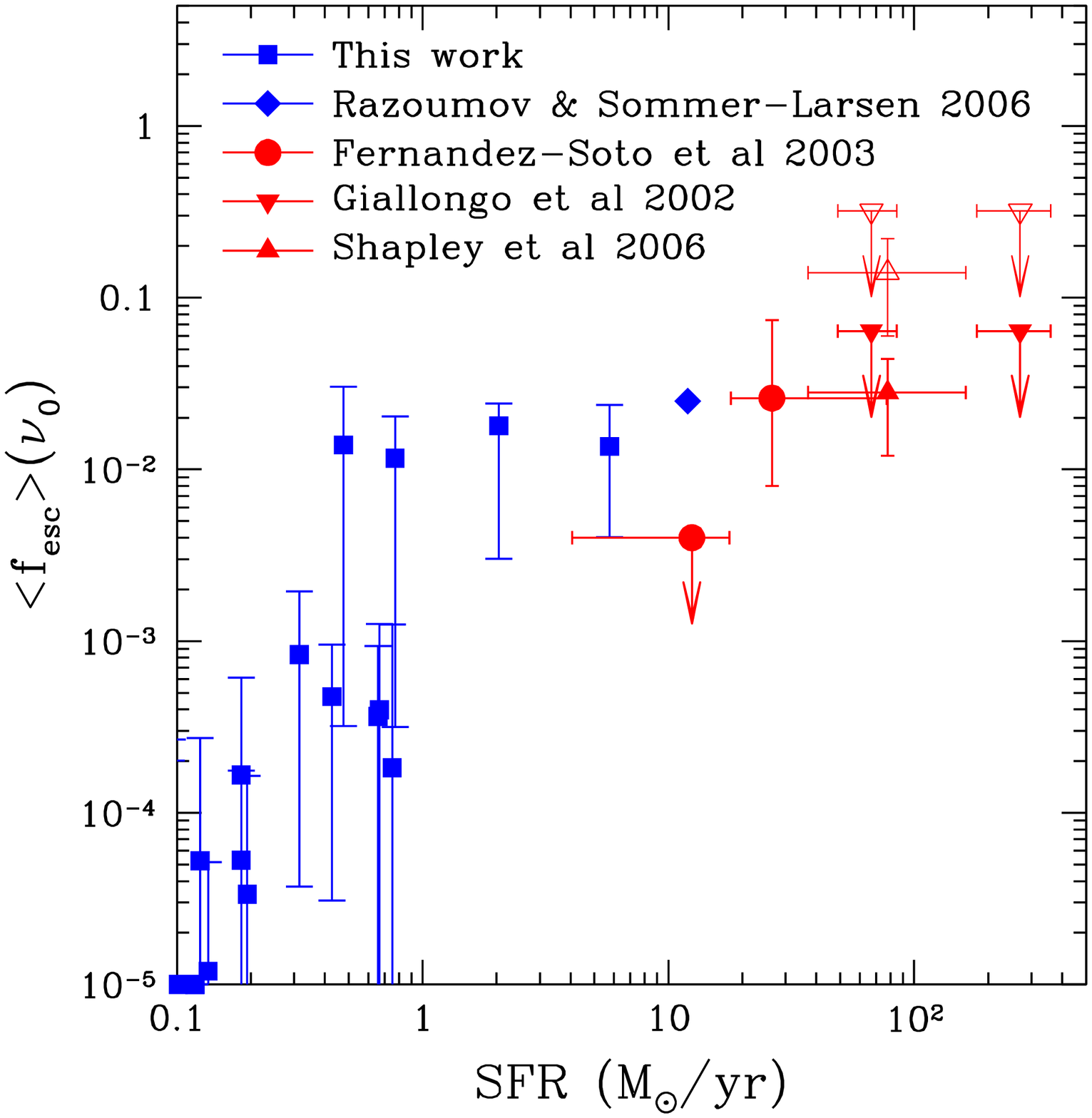}{f9.eps}}
\caption{A comparison of the angular averaged escape fraction at
  hydrogen ionization edge at $z=3$ from various
  determinations. Filled red/gray upward triangle and circles show the
  observational measurements from \citet{sspa06} and \citet{flc03}
  respectively, presented as average values for their data
  samples. Two filled red/gray downward triangles show the upper limits from
  \citet{gcdf02}. Both \citet{sspa06} are \citet{gcdf02} points
  are corrected 
  from the measured relative escape fraction to the absolute escape
  fraction, as explained in the text (open symbols show
  the original uncorrected measurements). The filled blue/black diamond
  is the simulation results of \citet{rs06} for the single
  galaxy they report the star formation rate for. Filled blue/black
  squares with error-bars show our results, similar to
  Figure~\ref{figFM}.}
\label{figFO}
\end{figure}

To summarize the current understanding of the escape fraction of
photons at the ionizing threshold of hydrogen, we show in
Figure~\ref{figFO} both simulation predictions and observational
constraints available at $z=3$. The simulation predictions are taken
from \citet{rs06} and our work. Observational constraints are from
\citet[][upper limits only]{gcdf02}, \citet{flc03}, and
\citet{sspa06}. The measurements of \citet{gcdf02} are for individual
galaxies with SFR taken from \citet{pksd98}. The measurement of
\citet{sspa06} is an average of 14 star-forming galaxies at $z\sim 3$.
The SFR is estimated based on the observed UV flux and corrected for
dust extinction for $\langle E(B-V)\rangle=0.11$. Both \citet{gcdf02}
and \citet{sspa06} measurements are corrected from the relative
measurements of the ratio $\fesc(912\AA)/\fesc(1500\AA)$ they report
to the absolute measurement of $\fesc(912\AA)$ by adopting a value of
$\fesc(1500\AA)=0.2$ from \citet{as00}. This correction factor is also
consistent with other observed estimates of reddening at $1500\AA$
\citep{pksd98,spa01,sspa06}.  The measurements of \citet{flc03} are
averaged over 14(13) high(low) luminosity galaxies found at $1.9 < z <
3.5$ in the Hubble Deep Field (the second one is actually an upper
limit).  The star formation rates are conversions of the observed UV
flux using the scaling relation from \citet{mpd98}, and are corrected
for dust extinction assuming $E(B-V)=0.1$ and the extinction law from
\citet{cabk00}. \citet{flc03} measure the absolute escape fraction, so
no correction is applied to their points.

There still exists a gap between simulations and observations in the
luminosity and SFR parameter space.  While our simulations model
modest-mass galaxies, observational measurements are reported mostly
for the brightest, $\ga L_{\ast}$ galaxies. Nevertheless, the
simulations are tantalizingly close to reaching a similar range of
luminosity and SFR covered by the observations. Both the observations
and our simulations indicate little (if any) trend in the {\it
absolute\/} escape fraction with the galaxy mass or star formation
rate, except for a rapid drop in $\fesc$ for smaller mass
galaxies. Our simulations indicate that this characteristic scale
corresponds to about SFR$\sim 1\,\Msun\,{\rm yr^{-1}}$ or $M_{\rm
tot}\sim\,10^{11}\Msun$, although an upper limit of $\fesc<0.4\%$ from
\citet{flc03} at SFR$\sim10\Msun\,{\rm yr^{-1}}$ may indicate that
simulations overestimate escape fractions in galaxies with
SFR$\la10\Msun\,{\rm yr^{-1}}$. The level of the disagreement (if any)
cannot yet be accurately deduced from our simulations, and, formally,
the \citet{flc03} upper limit is fully consistent with our results.
Note also that \citet{siana_etal07} in their recent study of the Lyman
continuum escape fractions in galaxies at $z\sim 1$ also find small
values ($\fesc\lesssim 0.1$), which is consistent with weak evolution
of escape fractions we find in our simulations.  The general agreement
between our results and observations is definitely encouraging and may
indicate that relative distribution of the young stars and UV
absorbing gas is modeled faithfully in our high-resolution AMR
simulations.

\begin{figure}[t]
\plotone{\figname{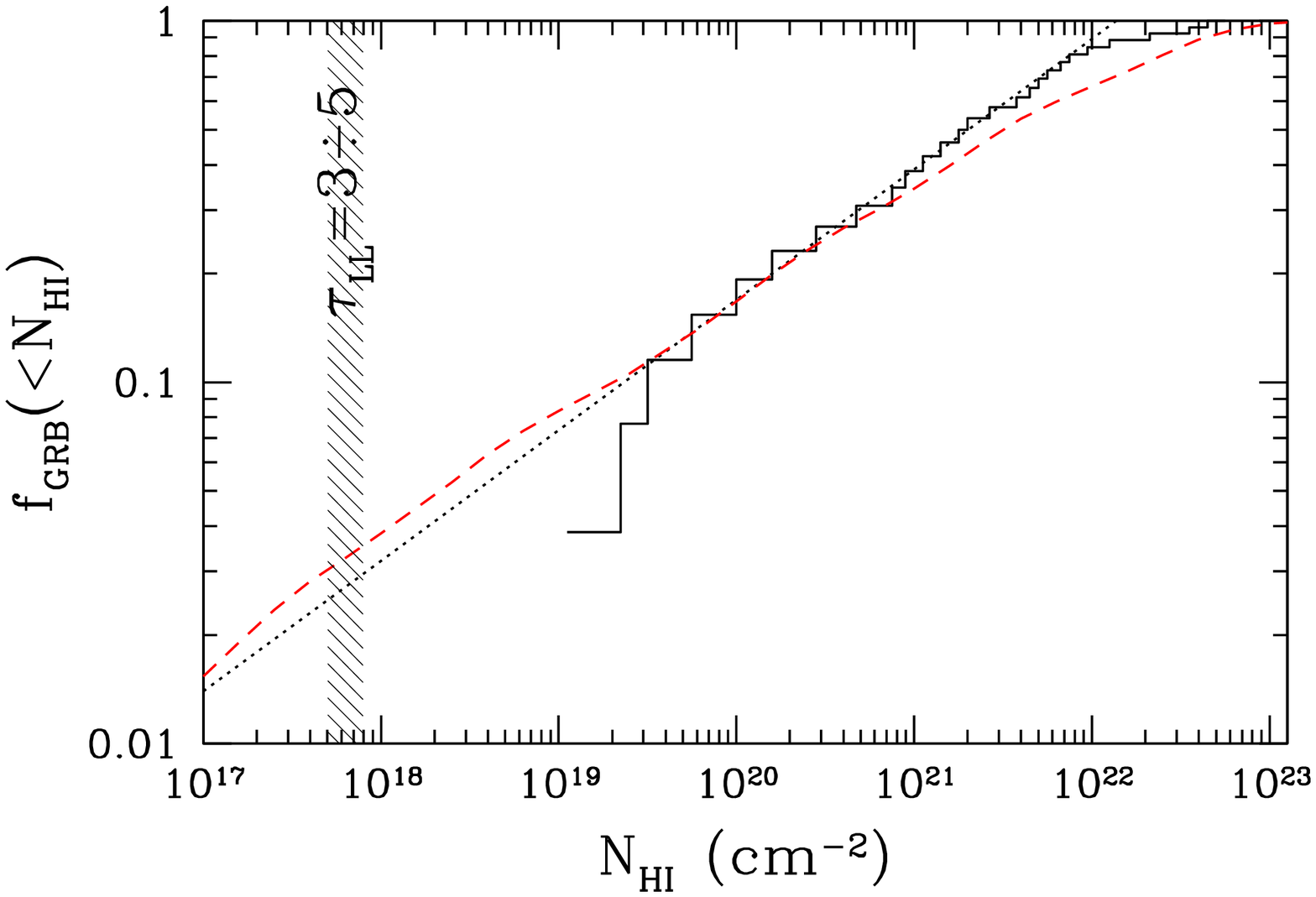}{f10.eps}}
\caption{A comparison of the cumulative distribution of $\HI$ column
densities toward ionizing sources from our simulation at $z=3$ (dashed
line) with that 
measured in the host galaxies of long duration $\gamma$-ray bursts at
$z>2$. The black solid line shows the observational data compiled by
\citet{velw04} and \citet{jflv06}, as well as the data for two
additional sources from \citet{sbpf06} and \citet{rstm07}. In two
cases (GRB 060124 and GRB 060607) only upper limits have been
measured. The dotted line is a power-law fit to the observational data
for the range of $\HI$ column densities between $10^{19.5}\dim{cm}^2$
and $10^{21.5}\dim{cm}^2$.}
\label{figGB}
\end{figure}
An interesting alternative way of constraining the escape fraction
observationally, independent of the observed UV light, is offered by
the observed distribution of $\HI$ column densities in the host
galaxies of long duration $\gamma$-ray bursts (GRBs; Chen, Prochaska,
\& Gnedin 2007, in preparation). Assuming that GRBs sample the  
distribution of young stars in an unbiased fashion, the fraction of
young stars located in ionized regions with small or negligible
attenuation of UV radiation should correspond to the fraction of GRBs
with $\HI$ column densities lower than some threshold value corresponding
to the transition between neutral and ionized hydrogen.

Figure~\ref{figGB} shows the observed distribution of $\HI$ column
densities in spectra of GRBs with confirmed redshifts of $z>2$,
compared to the corresponding distribution in our simulation at
$z=3$. The observed distribution is remarkably close to a power-law
for $10^{19.5}\dim{cm}^2 < N_{\HI} <
10^{21.5}\dim{cm}^2$. Extrapolating this power-law to lower column
densities of  $\approx (5-8)\times10^{17}\dim{cm}^2$
(or $\tau_{LL}=3-5$, appropriate for the boundary between the ionized and
neutral gas\footnote{The specific location of the ionization edge of a
galaxy depends on a number of factors and does not correspond to a
particular value of $\tau_{LL}$. A value of $\tau_{LL}=1$, which
corresponds to the suppression of ionizing background by only a factor
of 3, is certainly not enough to make the gas substantially neutral. A
value of $\tau_{LL}=10$ is, on the other hand, enough to make the gas
even at the mean cosmic density largely neutral. Thus, a value of
$\tau_{LL}$ around 3 to 5 should correspond to the approximate
location for the galaxy ionization edge.}) 
implies that about $2-3$\% of all GRBs are located in ionized regions
in which the UV radiation of their progenitors should have escaped
freely into the IGM, so that
$$
\fesc \approx f_{\rm GRB}(N_{\HI} < (5-8)\times10^{17}\dim{cm}^2) = 0.02-0.03.
$$
This value
is remarkably close to the escape fractions of $\approx 1-3\%$
measured in our simulations.  This test also provides strong support
for our interpretation of why the escape fractions are so low. 

Note that the $\HI$ column density distribution toward ionizing
sources (i.e. column densities from source locations to one virial
radius, as explained in \S 3) from our fiducial
simulation at $z=3$ is reasonably close to the observed distribution
of $N_{\HI}$ from GRBs for $N_{\HI}<3\times10^{21}\dim{cm}^2$. At
higher column density we do not expect a good agreement between our
simulation and the data, because the simulation does not incorporate
the physics of formation and self-shielding of molecular hydrogen
correctly, which becomes important in that regime. Thus, the
simulation is expected to overpredict $N_{\HI}$ column densities,
because in reality some of this $\HI$ is in molecular form.

While higher resolution simulations and deeper observations are needed
to bridge the remaining small gap in the probed star formation regime
between the data and the theory, our results suggest that average
escape fractions from bright galaxies at intermediate redshifts do not
depend strongly on galaxy properties or redshift. This is interesting
because this implies that porosity of the interstellar medium in
galaxies, and hence its opacity to ionizing radiation, does not
dramatically change at higher star formation rates, as expected in
some of the theoretical models \citep[e.g.,][]{co02}.  Note that
observed escape fractions at higher star formation rates are
consistent with being simply an extrapolation of the trend seen in
simulations at lower rates. If this is indeed the case, this implies
absence of a well-defined critical SFR above which the escape fraction
sharply increases to unity.

At the same time, the low values of escape fractions found
in our simulations suggest that high-redshift galaxies are quite
inefficient in emitting ionizing radiation produced by their young
stars into the intergalactic medium, with escape fractions decreasing
sharply for galaxies of $M_{\rm tot}\lesssim 10^{11}\Msun$ (or
SFR$\lesssim 1\Msun\dim{yr}^{-1}$).  This conclusion potentially has
important implications for the contribution from normal galaxies to
the early reionization of the universe and for the relative role of
galaxies and quasars in keeping the universe ionized at intermediate
redshifts.

\acknowledgements 

We thank Douglas Rudd for constructive and useful comments on the
draft of this paper.  This work was supported in part by the DOE and
the NASA grant NAG 5-10842 at Fermilab, by the NSF grants AST-0206216,
AST-0239759, and AST-0507596, and by the Kavli Institute for
Cosmological Physics at the University of Chicago. Supercomputer
simulations were run on the IBM P690 array at the National Center for
Supercomputing Applications and San Diego Supercomputing Center (under
grant AST-020018N) and the Sanssouci computing cluster at the
Astrophysikalisches Institut Potsdam. This work made extensive use of
the NASA Astrophysics Data System, {\tt arXiv.org} preprint server,
and the HEALPix\footnote{\tt http://healpix.jpl.nasa.gov}
\citep{ghbw05} package.

\appendix

\section{Escape fraction for ionizing radiation and the escape
  fraction at the ionization edge}

\begin{figure}[t]
\plotone{\figname{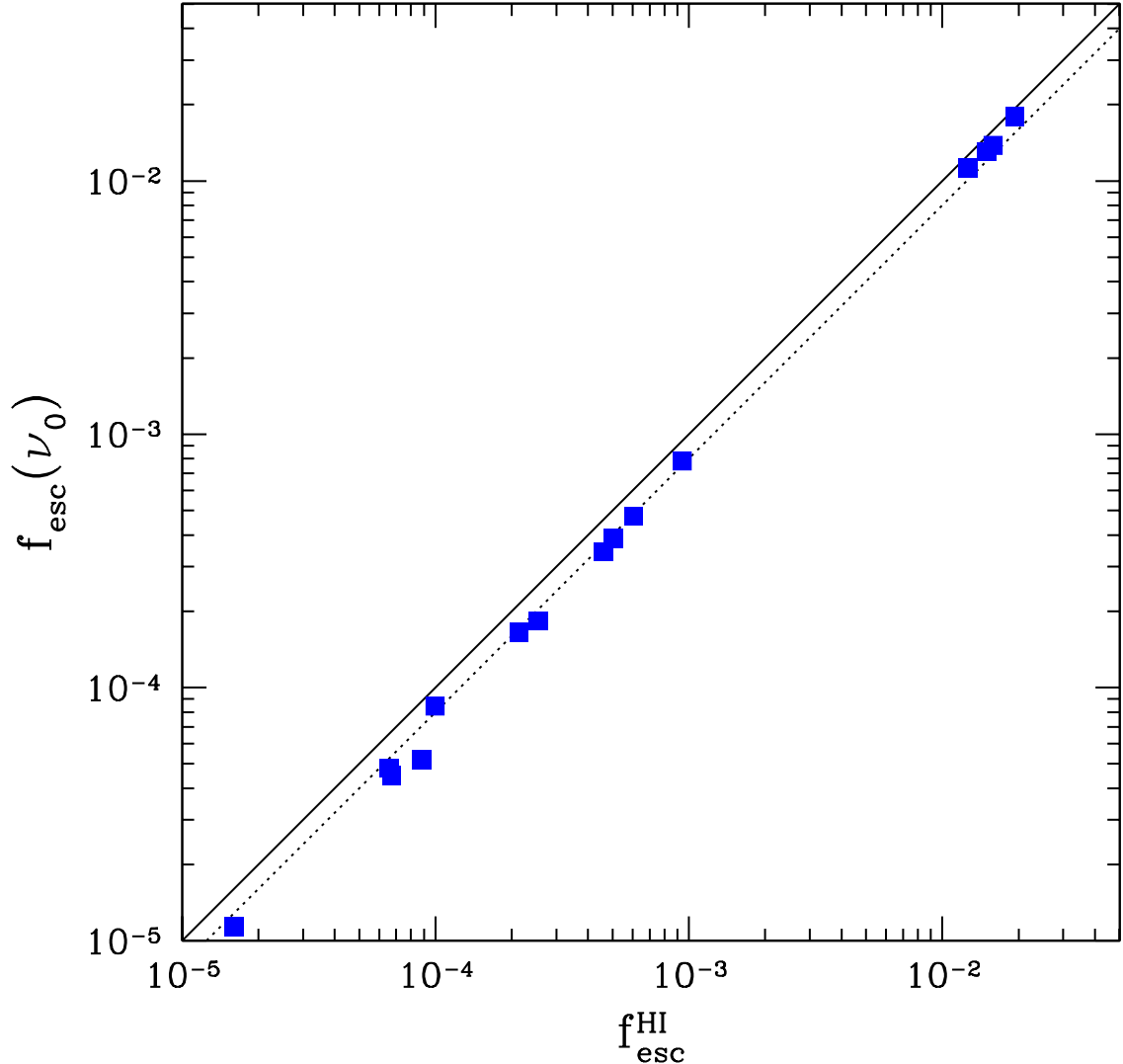}{f11.eps}}
\caption{A comparison of the angular averaged escape fraction for
  $\HI$ ionizing radiation (eq.\ \ref{eq:fion}) with the angular
  averaged escape fraction at the ionization edge (Lyman limit, the quantity
  usually measured in observational work), $h\nu_0=13.6\dim{eV}$. 
  Solid and dotted lines show the $\fesc^{\HI}=\fesc(\nu_0)$ and
  $\fesc^{\HI}=1.25\fesc(\nu_0)$ relations respectively.}
\label{figFF}
\end{figure}

In order to facilitate comparison between the theoretical and
observational results, in Figure\ \ref{figFF} we show a relationship
between the theoretically relevant escape fraction of ionizing
radiation (i.e.\ an integral quantity, as defined in eq.\
\ref{eq:fion}, since it determines the ionization state of the IGM)
and the escape fraction at the hydrogen ionization edge (Lyman limit),
which is usually measured in observational studies. We find a tight
correlation between the two quantities in the form
\begin{equation}
  \fesc^{\HI}=1.25\fesc(\nu_0),
  \label{eq:ff}
\end{equation}
which is helpful for relating observationally measured and
theoretically relevant quantities. We use this relation in Fig.\
\ref{figFO} when comparing observational and theoretical values on the
same plot.

\bibliographystyle{apj}
\bibliography{igm,gnedin,misc,dsh,grb}

\end{document}